\documentstyle[12lomcon,cite,amssymb,graphicx]{article}
\newcommand{\ds}{\displaystyle}
\renewcommand{\cdot}{\times}
\begin{document}

\begin{flushright}
DESY 06-090\\
June 2006\\
\end{flushright}

\title{RARE SEMILEPTONIC MESON DECAYS IN R-PARITY VIOLATING MSSM}

\author{A.~Ali \footnote{e-mail: ahmed.ali@desy.de}}

\address{Deutsches Electronen-Synchrotron, DESY, 22607 Hamburg, Germany}

\author{ A.V.~Borisov \footnote{e-mail: borisov@phys.msu.ru},
M.V.~Sidorova \footnote{e-mail: mvsid@rambler.ru}}

\address{Faculty of Physics, Moscow State University, 119992 Moscow, Russia}


\maketitle\abstracts{~We discuss rare meson decays $K^ +   \to \pi
^ -  \ell ^ +  \ell '^ + $ and $D^ +  \to K^ -  \ell ^ +  \ell '^
+$ ($\ell, \ell'=e, \mu$) in a supersymmetric extension of the
Standard Model with $R$-parity violation. Estimates of the
branching ratios for these decays are presented.}


In the standard model (SM), lepton $L$ and baryon $B$ number
conservation take place due to the accidental $U(1)_{L}\times
U(1)_B$ symmetry existing at the level of renormalizable
operators.  But, for many extensions of the SM, this is not the case.
The well known mechanism of lepton number (LN) violation is based
on the mixing of massive Majorana neutrinos predicted by various
Grand unified theories (GUTs) \cite{lan}. The Majorana mass term
violates LN by $\Delta L = \pm 2$ \cite{kay} and can lead to a
large number of LN violating processes. Among them, the most
sensitive to the LN violation are processes such as neutrinoless double
 beta decay $(A,Z) \to (A,Z+2)+e^-+e^-$ (for a recent review, see
 \cite{barab}), rare meson decays (see, e.g., \cite{abz,abs}) \vspace{-0.2cm}
\begin{equation}
\label{dec}
 M^{+}\to M^{'-}\ell^{+} \ell^{'+}~,
\end{equation}
and like-sign dilepton production in high-energy hadron-hadron and
lepton-hadron collisions (see, e.g., the papers and references
therein: $pp\to \ell^{\pm}\ell'^{\pm}X$ \cite{abz,abz1}, $e^ +  p
\to \bar \nu _e \ell ^ +  \ell '^ +  X$ \cite{abzh}), which have been
extensively studied in the literature.

There exists now convincing evidence for oscillations of solar,
atmospheric, reactor, and accelerator neutrinos \cite{pdg}. The
oscillations, i.e., periodic neutrino flavor changes, imply that
neutrinos have nonzero masses and they mix among each other: neutrinos
$\nu_{\ell}$ of specific flavors $\ell=e, \mu, \tau$ are the
coherent superposition of the neutrino mass eigen-states $\nu_N$
of masses $m_N$,
\vspace{-0.2cm}
\begin{equation}
\label{nu} \nu _\ell   = \sum\nolimits_N {U_{\ell N} \nu _N }.
\vspace{-0.2cm}
\end{equation}
Here the coefficients $U_{\ell N}$ are elements of the unitary
leptonic mixing matrix -- the PMNS matrix~\cite{PMNS}.

Neutrino flavor changes imply lepton family number $L_{\ell}$
nonconservation admissible for neutrinos of both types, Dirac and
Majorana, but for the Dirac neutrinos, in contrast to the Majorana ones,
the total lepton number $L = \sum\nolimits_{\ell}{L_\ell  }$ is
conserved. The nature of the neutrino mass is one of the
main unsolved problems in particle physics. However, oscillation
experiments can not distinguish between the two types of neutrinos.

In Refs. \cite{abz,abs} we investigated the rare decays (\ref{dec}) of
the pseudoscalar mesons $M=K, D, D_s, B$, mediated by light
($m_N  \ll m_\ell  ,\;m_{\ell '} $ ) and heavy ($m_N  \gg m_M $)
Majorana neutrinos. It was shown that the present direct
experimental bounds on the branching ratios (BRs) are too weak to
set reasonable limits on the effective Majorana masses. Taking
into account the limits on lepton mixing and neutrino masses
obtained from the precision electroweak measurements, neutrino
oscillations, cosmological data and searches of the neutrinoless
double beta decay, we have derived the indirect upper bounds on
the BRs that are greatly more stringent than the direct ones.

In this report, we investigate another mechanism of the  $\Delta L =2$ rare
decays (\ref{dec}) based on  $R$-parity violating supersymmetry
(SUSY) (for a review see~\cite{bar}). We recall that $R$-parity
is defined as $R=(-1)^{3(B-L)+2S}$, where $S$, $L$, and $B$
are the spin, the lepton and baryon numbers, respectively.
In the minimal supersymmetric extension of the SM (MSSM),
$R$-parity conservation is imposed to prevent the $L$ and $B$
violation; it also leads to  the production of superpartners in pairs and
ensures the stability of the lightest superparticle. However, neither gauge
invariance nor supersymmetry require $R$-parity conservation.
There are many generalizations of the MSSM with explicitly or
spontaneously broken $R$-symmetry \cite{bar}. We consider a SUSY
theory with the minimal particle content of the MSSM and explicit
$R$-parity violation ($\not\!\! R$MSSM).

The most general form for the $R$-parity and lepton number
violating part of the superpotential is given by \cite{bar,por}
\vspace{-0.2cm}
\begin{equation}
\label{rpv}
W_{\not R}  = \varepsilon _{\alpha\beta } \left(
{\frac{1} {2}\lambda _{ijk} L_i^\alpha  L_j^\beta  \bar E_k  +
\lambda '_{ijk} L_i^\alpha  Q_j^\beta  \bar D_k  + \epsilon _i
L_i^\alpha H_u^\beta  } \right).
\vspace{-0.2cm}
\end{equation}
Here $i, j, k =1, 2, 3$ are generation indices, $L$ and $Q$ are
$SU(2)$ doublets of left-handed lepton and quark superfields
($\alpha, \beta = 1, 2$ are isospinor indices), $\bar{E}$ and
$\bar{D}$ are singlets of right-handed superfields of leptons and
down quarks, respectively; $H_u$ is a doublet Higgs superfield
(with hypercharge $Y=1$); $\lambda _{ijk} = - \lambda _{jik}
,~\lambda '_{ijk}$ and $\epsilon _i $ are constants.

In the superpotential (\ref{rpv}) the trilinear ($\propto \lambda
,~\lambda '$) and bilinear ($\propto \epsilon $) terms are present.
In this work, we assume that the bilinear terms are absent at tree
level ($\epsilon =0$). They will be generated by quantum
corrections \cite{bar}, but it is expected that the phenomenology
will still be dominated by the tree-level trilinear terms.


At first we consider the rare decay $ K^{+} (P)\to \pi^{-}(P^{'})+
\ell^{+}(p)+ \ell^{'+}(p^{'})$ in the $\not\!\! R$MSSM (a rough
estimate of the width of the decay $\mbox{B}(K^{+}\to \pi ^{-}\mu
^{+}\mu ^{+})$ in the same theory was obtained in \cite{ls}). The
leading order amplitude of the process is described by three types
of diagrams shown in Fig. 1.
\begin{figure}
\begin{center}
\includegraphics[scale=0.5]{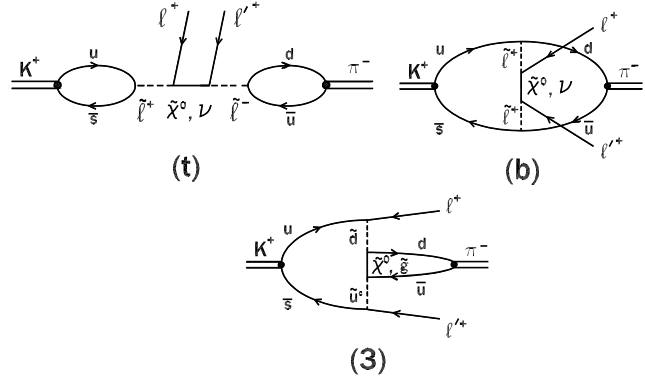}
\vspace{-0.5cm} \caption{Feynman diagrams for the  decay  $
K^{+}\to \pi^{-}+ \ell^{+}+ \ell^{'+}$ mediated by Majorana
neutrinos $\nu$, neutralinos $\tilde \chi ^0 $, gluinos $\tilde g$
with $\tilde f$ being the scalar superpartners of the
corresponding fermions $f=\ell, u, d$ (leptons and quarks). Bold
vertices correspond to Bethe--Salpeter amplitudes for mesons as
bound states of a quark and an antiquark; {\bf (t), (b)} and {\bf (3)}
stand for tree, box and simply the third 
kind of diagrams, respectively.  
There are also crossed
diagrams with interchanged lepton lines.}
\end{center}
\vspace{-0.5cm}
\end{figure}

The hadronic parts of the decay amplitude are calculated with the use
of a model for the Bethe--Salpeter amplitudes for mesons \cite{est},
\vspace{-0.2cm}
\begin{equation}
\label{bsq} \chi _{P}(q) = \gamma^5 (1  - \delta_M
{\not\!{P}})\varphi_{P}(q),
\vspace{-0.2cm}
\end{equation}
where $\delta_M = (m_1+m_2)/m_M^2$, $m_M$ is the mass of the meson
composed of a quark $q_1$ and an antiquark $\bar{q}_2$ with the
current masses $m_1$ and $m_2$, $P=p_1+p_2$ is the total
4-momentum of the meson, $q=(p_1-p_2)/2$ is the quark-antiquark
relative 4-momentum; $\varphi_{P} (q)$ is the model-dependent
scalar function.

For all mesons in question, $m_M  \ll m_{SUSY}$, where $m_{SUSY}
\gtrsim 100~\mbox{GeV}$ is the common mass scale of superpartners,
and for heavy Majorana neutrinos, $m_N  \gg m_M $ (the
contribution of light neutrinos is strongly suppressed by
phenomenology \cite{abz,abs}), we can neglect momentum dependence
in the propagators (see Fig.~1) and use the effective low-energy
current-current interaction. In this approximation the decay
amplitude does not depend on the specific form of the functions
$\varphi_{P} (q)$ (see Eq. (\ref{bsq})) and is expressed through
the known decay constants of the mesons, $f_M$, as
\vspace{-0.2cm}
\[
f_M =4\sqrt{N_{c}}~\delta _{M}\left( 2\pi \right)
^{-4}\int d^{4}q\, \varphi_P\left(q\right), \vspace{-0.2cm}
\]
where $N_c = 3$ is the number of colors.

For the total width of the decay we obtain
\vspace{-0.2cm}
\begin{eqnarray}
\label{GK} &\ds  \Gamma (K^ +   \to \pi ^ -  \ell ^ +  \ell '^ + )
= \bigl( {1 - \frac{1} {2}\delta _{\ell \ell '} }
\bigr)\frac{{f_K^2 f_\pi ^2 m_K^3 }} {{2^{12} \pi ^3 \delta _K^2
\delta _\pi ^2 }}\Phi _{\ell \ell '}   \nonumber \\ &\ds \times
\biggl| {\sum\limits_{i,j,k,k',N} {(\lambda _{ik\ell }^ *  } }
\biggr.\lambda _{jk'\ell '}^ *   + \lambda _{ik\ell '}^ *  \lambda
_{jk'\ell }^ *  )\frac{{\lambda '_{k12} \lambda '_{k'11} U_{iN}
U_{jN} }} {{m_{\tilde \ell _{Lk} }^2 m_{\tilde \ell _{Lk'} }^2 m_N
}}\bigl( {1 - \frac{1} {{2N_c }}} \bigr)  \nonumber \\ &\ds +\,
(\lambda '_{\ell 11} \lambda '_{\ell '12}  + \lambda '_{\ell '11}
\lambda '_{\ell 12} )\biggr[ {g_2^2 \sum\limits_{\delta  = 1}^4
{\frac{1} {{m_{\tilde \chi _\delta  } }}} \biggl(
{2\frac{{\epsilon _{L\delta }^ *  (\ell )\epsilon _{L\delta }^
*  (\ell ')}} {{m_{\tilde \ell L}^2 m_{\tilde \ell 'L}^2 }}}
\biggr.\bigl( {1 - \frac{1} {{2N_c }}} \bigr)} \biggr. \nonumber\\
&\ds - \bigl. {\frac{1} {{N_c }}\frac{{\epsilon _{R\delta }
(d)\epsilon _{L\delta }^
*  (u)}} {{m_{\tilde d_R }^2 m_{\tilde u_L }^2 }}} \biggr) +
\biggl. {\biggl. {\frac{{4g_3^2 }} {{N_c^2 }}\frac{1} {{m_{\tilde
d_R }^2 m_{\tilde u_L }^2 m_{\tilde g} }}} \biggr]} \biggr|^2.&
\vspace{-0.3cm}
\end{eqnarray}
Here $\Phi_{\ell \ell'}$ is the reduced phase space integral
($z=(P-P')^2/m_K^2$):
\vspace{-0.2cm}
\[
 \Phi _{\ell \ell '}  = \int\nolimits_{l_ +  }^{h_ -  } {dz}
\left( {1 - \frac{{l_ +   + l_ -  }} {{2z}}} \right)\left[ {(h_ +
- z)(h_ -   - z)(l_ +   - z)(l_ -   - z)} \right]^{1/2},
\vspace{-0.3cm}
\]
and the various parameters are defined as follows:
\vspace{-0.2cm}
\begin{eqnarray*}
& h_ \pm   = \left(1 \pm m_\pi/m_K \right)^2 ,\quad l_ \pm =
\left[ (m_\ell \pm m_{\ell '})/m_K  \right]^2;\\ &
\epsilon_{L\delta}(\psi)=-T_3(\psi)N_{\delta
2}+\tan{\theta_W}(T_3(\psi)-Q(\psi))N_{\delta 1},\\ &
\epsilon_{R\delta}(\psi)=Q(\psi)\tan{\theta_W}N_{\delta 1},
\vspace{-0.3cm}
\end{eqnarray*}
where $Q(\psi)$ and $T_3(\psi)$ are the electric charge and the third
component of the weak isospin for the field $\psi$, respectively, and
$N_{\delta\sigma}$ is the $4\times 4$ neutralino mixing matrix.
For the numerical estimates of the branching ratios, $\mbox{B}_{\ell \ell '} =
\Gamma \left( {M^ +   \to M'^ - \ell ^ +  \ell '^ +  }
\right)/\Gamma _{\rm total}$, we have used the known values for
the couplings, decay constants, meson, lepton and current quark
masses \cite{pdg,abs}, and a typical set of the matrix elements
$N_{\delta 1}, N_{\delta 2}$ from Ref. \cite{all}. In addition, we
have taken all the masses of superpartners to be equal with a common value
$m_{SUSY}$. Taking into account the present bounds on the
effective inverse Majorana masses \cite{abs}, we find that the
main contribution to the decay width  comes from the exchange by
neutralinos and gluinos (see Fig.~1). The results of the
calculations with the use of Eq. (\ref{GK}) for the decays
 $K^+ \to \pi^- \ell^+\ell^+$, and an analogous formula
for the decays $D^+ \to K^- \ell ^+ \ell '^+$, are shown in the
fourth column of Table 1 (here $ m_{200}  = m_{SUSY} /(200
\mbox{GeV})$). In the second and third columns of this table, the present
direct experimental upper bounds on the BRs \cite{pdg} and the indirect
bounds for the Majorana mechanism of the rare decays \cite{abs}
are shown, respectively. Our result for the $\mbox{B}(K^{+}\to \pi
^{-}\mu ^{+}\mu ^{+})$ is in agreement with a rough estimate of
Ref. \cite{ls}.

To calculate the upper bounds on the BRs in the $\not\!\! R$MSSM,
we take $m_{200}  = 1$ and $|\lambda '_{ijk} \lambda
'_{i'j'k'}|\lesssim 10^{-3}$ \cite{tah}. It yields
\vspace{-0.2cm}
\[
\mbox{B}\left( {K^ +   \to \pi ^ -  \ell ^ +  \ell '^ +  } \right)
\lesssim 10^{ - 23} ,\;\mbox{B}\left( {D^ +  \to K^ -  \ell ^ +
\ell '^ +  } \right) \lesssim 10^{ - 24}.
\vspace{-0.2cm}
\]
These estimates are much smaller than the corresponding direct
experimental bounds but are close (except for the $ee$ decay mode)
to the indirect bounds based on the Majorana mechanism of the
decays (see Table 1).

\vspace{-0.4cm}

\begin{table}[htb!]
\caption{The branching ratios $\mbox{B}_{_{\ell \ell ^{\prime }}}$
for the rare meson decays $M^{+}\to M^{\prime
-}\ell^{+}\ell^{\prime +}.$} {
\begin{center}
\begin{tabular}{|c|c|c|c|}
\hline Rare decay&Exp. upper  &Ind. bound&$\mbox{B}_{\ell
\ell^{\prime }}\cdot m_{200}^{10}$\\ &bound on $\mbox{B}_{\ell
\ell '} $& on $\mbox{B}_{\ell \ell '} (\nu_{M}\mbox{SM})
$&(${\not\!R}\mbox{MSSM}$)\\ \hline\hline $K^{+}\to \pi
^{-}e^{+}e^{+}$ &$6.4\cdot 10^{-10}$&$5.9\cdot 10^{-32}$&$5.7\cdot
10^{-17}|\lambda'_{111}\lambda'_{112}|^2$
\\ \hline

$K^{+}\to \pi ^{-}\mu ^{+}\mu ^{+}$ & $3.0\cdot 10^{-9}$&$1.1\cdot
10^{-24}$&$2.0\cdot 10^{-17}|\lambda'_{211}\lambda'_{212}|^2$
\\ \hline

$K^{+}\to \pi ^{-}e^{+}\mu ^{+}$ & $5.0\cdot 10^{-10}$&$5.1\cdot
10^{-24}$& $1.9\cdot
10^{-17}|\lambda'_{111}\lambda'_{212}+\lambda'_{211}\lambda'_{112}|^2$
\\ \hline\hline
$D^{+}\to K ^{-}e^{+}e^{+}$ & $1.2\cdot 10^{-4}$&$1.5\cdot
10^{-31}$&$1.0\cdot
10^{-18}|\lambda'_{122}\lambda'_{111}+0.45\lambda'_{121}\lambda'_{112}|^2$
\\ \hline

$D^{+}\to K ^{-}\mu ^{+}\mu ^{+}$ & $1.3\cdot 10^{-5}$&$8.9\cdot
10^{-24}$&$9.6\cdot
10^{-19}|\lambda'_{222}\lambda'_{211}+0.45\lambda'_{221}\lambda'_{212}|^2$
\\ \hline
$D^{+}\to K ^{-}e^{+}\mu ^{+}$ &$1.3\cdot 10^{-4}$&$2.1\cdot
10^{-23}$&$4.9\cdot
10^{-19}|(\lambda'_{122}\lambda'_{211}+\lambda'_{222}\lambda'_{111})$\\
&&&$+0.45(\lambda'_{121}\lambda'_{212}+\lambda'_{221}\lambda'_{112})|^2$
\\ \hline
\end{tabular}
\end{center}}
\end{table}

\section*{Acknowledgments}

We thank K. V. Stepanyantz, Kamal Ahmed, Farida Tahir and Werner
Porod for helpful discussions.


\end{document}